\documentclass[aps,prb,superscriptaddress,showpacs,twocolumn]{revtex4}
\usepackage{graphicx}
\newcommand{\Co} {{\mbox{Co${}_7$(Te${}$O${}_{3}$)${}_4$Br${}_6$}}}
\newcommand{\astar} {\mbox{$a^\ast$}}
\newcommand{\cstar} {\mbox{$c^\ast$}}
\begin{document}
\title{Ferromagnetism in \Co: A byproduct of complex antiferromagnetic order and single-ion anisotropy}


\author{M. Prester} 
\affiliation{Institute of Physics, P.O.B.304, HR-10 000, Zagreb, Croatia}
\author{I. \v Zivkovi\'c}
\affiliation{Institute of Physics, P.O.B.304, HR-10 000, Zagreb, Croatia}
\author{O. Zaharko}
\affiliation{Laboratory for Neutron Scattering, ETHZ \& PSI, CH-5232, Villigen, Switzerland} 
\author{D. Paji\'c}
\affiliation{Department of Physics, Faculty of Science, Bijeni\v{c}ka c.32 , HR-10 000 Zagreb, Croatia} 
\author{P. Tregenna-Piggott}
\affiliation{Laboratory for Neutron Scattering, ETHZ \& PSI, CH-5232, Villigen, Switzerland}
\author{H. Berger}
\affiliation{Institute of Physics of Complex Matter, EPFL, 1015 Lausanne, Switzerland}

\date{\today}

\begin{abstract}
Pronounced anisotropy of magnetic properties and complex magnetic order of a new oxi-halide compound $\Co$ has been investigated by powder and single crystal neutron diffraction, magnetization and ac susceptibility techniques. Anisotropy of susceptibility extends far into the paramagnetic temperature range. A principal source of anisotropy are anisotropic properties of the involved octahedrally coordinated single Co$^{2+}$ ions, as confirmed by angular-overlap-model calculations presented in this work.
Incommensurate antiferromagnetic order sets in at $T_{N}$=34 K. Propagation vector is strongly temperature dependent reaching ${\bf{k}_1}$=(0.9458(6), 0, 0.6026(5)) at 30 K. A transition to a ferrimagnetic structure with ${\bf{k}_2}$=0 takes place at $T_{C}$=27 K. 
Magnetically ordered phase is characterized by very unusual anisotropy as well: while $M-H$ scans along $b$-axis reveals spectacularly rectangular but otherwise standard ferromagnetic hysteresis loops, $M-H$ studies along other two principal axes are perfectly reversible, revealing very sharp spin flop (or spin flip) transitions, like in a standard antiferromagnet (or metamagnet). \\
Altogether, the observed magnetic phenomenology is interpreted as an evidence of competing magnetic interactions permeating the system, first of all of the single ion anisotropy energy and the exchange interactions. Different coordinations of the Co$^{2+}$-ions involved in the low-symmetry C2/c structure of  $\Co$ render the exchange-interaction network very complex by itself. Temperature dependent changes in the magnetic structure, together with an abrupt emergence of a ferromagnetic component, are ascribed to continual spin reorientations described by a multi-component, but yet unknown, spin Hamiltonian.

\end{abstract}


\maketitle

%
%
%
%

\section{Introduction}

Enormous diversity of magnetic phenomena relies on equal diversity of interactions permeating real magnetic materials. In cases of one interaction dominating by far over the others (typically valid for exchange interaction) an elementary insight into magnetism of such a system can indeed be acquired on basis of a simple, one-interaction Hamiltonian. However, any profound knowledge of magnetism of even such a simple system implies taking into account other interactions present in the system, whatever weak they could be. For example, basic static properties (like magnetization, susceptibility, magnetic structure...) of long-range ordered ferromagnets and antiferromagnets, especially if they rely on strong exchange interaction, can be well-interpreted within one-interaction models. Understanding dynamic features of the same systems, like their spin excitation spectra for example, requires however at least magnetocrystalline anisotropy (thus, in turn, several inevitably involved specific interactions) to be included as well. The most interesting situation, for fundamental research as well for applications, arises in cases involving presence of magnetic interactions competing each other in size and/or in sign while residing on energy scale of thermal excitations within the usual experimental window $2-300$ K. Antiferromagnetic exchange competing with single ion anisotropy can, for example, render the certainty of long-range ordering questionable, as shown decades ago for archetypal antiferromagnet~\cite{Moriya1960PRB} NiF$_2$. In a more recently developed general framework of Quantum Magnetism~\cite{Lhuillier2002} the subject of competing interactions is recognized as a key ingredient introducing the quantum phase transition point into the respective phase diagrams. In another aspect, competing interactions are responsible for complex/incommensurate magnetic structures and emergence of ferromagnetism in the ground states of many nominally antiferromagnetic systems~\cite{Bogdanov2002}. In this category there is a particularly interesting group of systems revealing spin reorientations taking place within the system-specific temperature intervals. The prominent examples are the rare-earth elements Dysprosium (Dy) and Terbium (Tb)~\cite{Nagamiya1967}, transition-metals sesquioxides~\cite{Artman65} (notably, hematite~\cite{Besser1967}, $\alpha$-Fe$_2$O$_3$) and orthoferrites~\cite{Shapiro1974}. From applicative side there is renewed interest for these systems, particularly for orthoferrites~\cite{Kimel2004}: spins in these systems are subject to combined effect of antiferromagnetic exchange and magnetocrystalline anisotropy which enables ultrafast spin manipulation~\cite{Kimel2004}. Ultrafast dynamics is a key issue for exchange-bias devices.

In this article we present magnetic structure and properties of a recently discovered~\cite{Becker2006} magnetic system, $\Co$, which, as shown herewith, reveals remarkable manifestations of competing magnetic interactions. In its ground state there is a complex noncolinear long-range magnetic order while pronounced magnetic anisotropy characterizes both paramagnetic and magnetically ordered phase. In the temperature range $27K-34K$ the incipient incommensurate magnetic order continuously changes by cooling, ending up at $27K$ by an abruptly intruding ferromagnetic component along one crystallographic axis. In the spin reorientation, underlying emergence of ferromagnetism, antiferromagnetic backbone remains conserved: we show that the key hallmark of antiferromagnetic order, spin flop/flip transition, keeps characterizing magnetism of $\Co$ in measurements along other two principal axes. The results are interpreted by invoking competition of exchange interactions primarily with single ion anisotropy energy but also with several other possible sources of magnetocrystalline anisotropy.

The article is organized as follows. The involved experimental techniques are introduced in Section \ref{Details}. In Section \ref{Results} results of neutron diffraction, dc/ac susceptibility as well as of magnetic hysteresis/M-H studies are presented. Results of theoretical modeling of paramagnetic susceptibility and of the ground state properties, based on angular-overlap-model calculations, are also presented in Section \ref{Results}. Both experimental and theoretical results provide firm evidence of exchange interactions and single ion anisotropy energy ruling magnetic properties of $\Co$. In Section \ref{Discussion} these results are discussed in a more general framework of competing interactions of various sorts, knowledge of which was accumulating during the decades. In Section \ref{Conclusion} appropriate conclusions have been presented.

%
%
%
%

\section{Experimental details}
\label{Details}

The single crystals of $\Co$ were synthesized via chemical vapor transport reactions. The details of the synthesis can be found elsewhere~\cite{Becker2006}. The single crystals grow in platelet geometry and the typical samples used in the magnetization/susceptibility studies had approximate dimensions $4.0 \times 2.0 \times 0.1~mm ^3$. The plane of the platelet-like samples corresponds to the crystallographic $bc$-plane. 

Neutron powder diffraction data have been collected in the temperature range 3 K - 45 K on the DMC instrument at SINQ, Paul Scherrer Institute, Villigen, Switzerland, with neutron wavelength of 2.453~\AA ~and 4.2~\AA. Incommensurate (ICM) wave vector has been determined on the single crystal diffractometer TriCS at SINQ with neutron wavelength of 2.32~\AA ~using the area detector.

Magnetization/susceptibility measurements were performed on oriented single crystals in applied magnetic fields directed parallel to the crystallographic $\astar$-, $b$-, and $c$-axis. Samples were oriented by the use of X-ray diffractometer. The choice of $\astar$-axis sample orientation, instead of the preferable $a$-axis one, is imposed by sample morphology. For DC magnetization studies up to 5.5 T a Quantum Design superconducting quantum interference device (SQUID) magnetometer was used, covering the temperature range 2 -- 300 K. AC susceptibility studies were performed using a CryoBIND ac susceptibility system employing measuring ac field of 3 Oe and frequency of 430 Hz.

%
%
%
%

\section{Results}
\label{Results}

\subsection{Neutron diffraction}

Refinement of the neutron powder pattern collected at 45 K confirmed that the sample is  a single phase with a monoclinic \emph{C2/c} space group and unit cell parameters $a=20.590(2)$~\AA , $b=8.4998(7)$~\AA , $c=14.631(1)$~\AA , $\beta = 125.202(6) ^\circ$, in close agreement with the original structural work~\cite{Becker2006}. As pointed out therein the crystal structure can be described as layered in the $bc$-plane. The layers are built of networks comprising three types of Co$^{2+}$-ion distorted octahedra ([Co(1)O$_4$Br$_2$], [Co(2)O$_4$Br$_2$], [Co(3)O$_4$Br$_2$]) and [TeO$_3$E] tetrahedra, while the layers are interconnected along the $a$-axis by the fourth Co$^{2+}$-ion octahedron type, [Co(4)O$_2$Br$_4$].  

\begin{figure}
\includegraphics[width=0.4\textwidth]{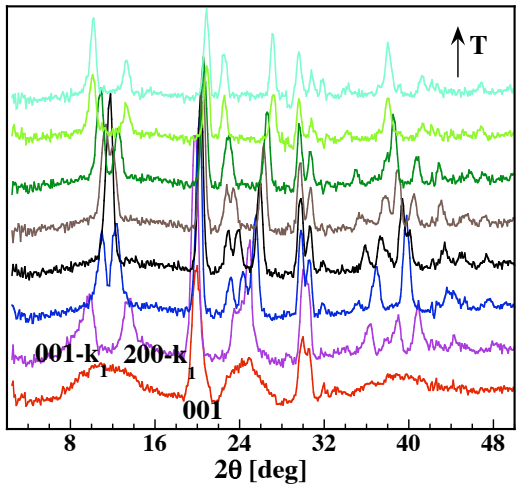}
\caption{(Color online) Difference neutron powder diffraction $T - 45$ K of $\Co$ with $T$ in the range 25 -- 32 K (DMC instrument, $\lambda = 4.2$~\AA).}
\label{neutron_peaks}
\end{figure}

\begin{figure}
\includegraphics[width=0.4\textwidth]{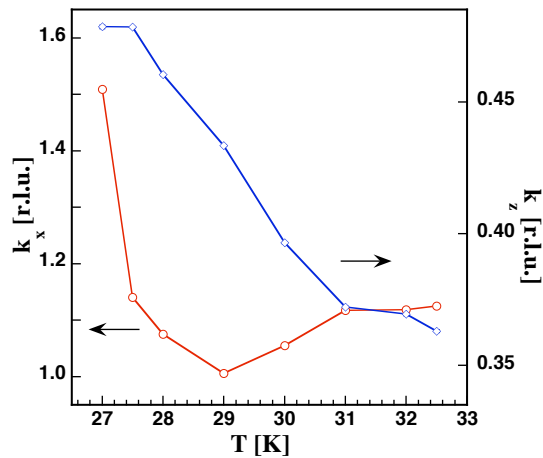}
\caption{(Color online) Temperature dependence of the ICM magnetic vector components refined from the DMC data, $\lambda =2.453$~\AA.}
\label{neutron_kxkz}
\end{figure}

The collected neutron powder diffraction patterns confirmed the existence of two sequential magnetic orderings, as claimed by the original work on basis of magnetic susceptibility studies~\cite{Becker2006}. Below $T_N=34$ K weak magnetic peaks occur. They cannot be indexed as simple multiples of the crystal unit cell revealing an incommensurate wave vector. Moreover, the incommensurability appears in two directions, $\astar$ and $\cstar$, which has been clarified by a single crystal study using the 2D detector of TriCS. The position of the ICM peaks is strongly temperature dependent and in Fig.~\ref{neutron_peaks} we show the difference neutron powder diffraction patterns T-45 K in the range T= 25 -- 32 K. Each pattern presented in Fig.~\ref{neutron_peaks} has been analyzed using the profile match option of the Fullprof program\cite{Rodriguez1993} and the two ICM components $(k_x, 0, k_z)$ of the wave vector has been refined. The temperature dependence of $k_x$ and $k_z$ is shown in Fig.~\ref{neutron_kxkz}.


At 27.5 K a second set of additional strong magnetic peaks appears, coexisting in a short temperature interval with the ICM peaks. Further temperature lowering leads to weakening of the ICM peaks and their transformation into diffuse scattering, as shown in the bottom pattern of Fig.~\ref{neutron_peaks}. Below 26 K only the second set of magnetic peaks remains.
%
\begin{figure}
\includegraphics[width=0.45\textwidth]{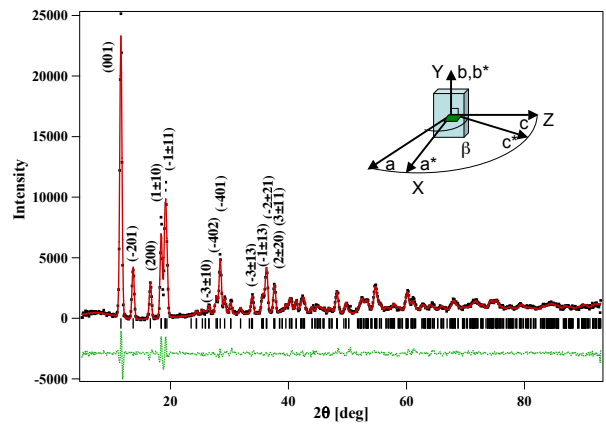}
\caption{(Color online) Observed 5~K - 45~K magnetic difference pattern, calculated and difference patterns of ${\Co}$ denoted by crosses, red solid, and green dotted lines, respectively. Inset: The choice of the $XYZ=a^*bc$ orthogonal system and morphology of the single crystal used.}
\label{k0}
\end{figure}

This set can be indexed with the wave vector {\bf k$_2$} = 0. Some of the new magnetic peaks (i.e. (200), (110)) overlap with the nuclear reflections implying ferromagnetic contribution, others (i.e. (001), (-201)) appear at the positions extinct in the paramagnetic pattern, as expected from an antiferromagnetic component. The systematic extinctions observed in the 5 K - 45 K magnetic difference pattern reveal that the C-translation and the glide plane $c$ not combined with the time reversal are retained in magnetic symmetry ($hkl: h + k = 2n$ and $h0l: l \neq 2n$).

Representation analysis implemented in the Fullprof program~\cite{Rodriguez1993} has been used to determine the {\bf k$_2$} = 0 magnetic structure. The Fourier coefficients describing possible spin configurations can be written as linear combinations of irreducible representations (IR) of the wave vector group (little group). The magnetic representations for the 4a and 8f sites, occupied by Co(4) and Co(1-3), can be decomposed in IR's: \\
$\Gamma$(4a)= 3 $\Gamma_1$ + 3 $\Gamma_3$\\
$\Gamma$(8f)= 3 $\Gamma_1$ + 3 $\Gamma_2$ + 3 $\Gamma_3$ + 3 $\Gamma_4$
 \\
Note that there are four ions of the Co(4) set and eight ions in each of the Co(1), Co(2) and Co(3) sets.
In the $\Gamma_1$ and $\Gamma_3$ representations magnetic moments of the Co(1)-Co(4) sets may attain independent values and directions, while the moments of the ions within the same site are constrained by the symmetry relations presented in Table~\ref{table1}.
So all together there are twelve independent parameters ($m_x$, $m_y$ and $m_z$ of four sets of Co$^{2+}$ ions) in $\Gamma_1$ and $\Gamma_3$.
In $\Gamma_2$, $\Gamma_4$ IR's the magnetic moments of the Co(4) set must be zero.
%
%
\begin{table}
\caption{
Irreducible representations of the wave vector group for ${\bf{k}_2}$= 0 in the space group $C 2/c$. The notation Co($ij$) is used with the
index $i$=1,...,4 labeling the set, and the index $j$=1,...,4 labeling the ions within the site. The ions of the same site have coordinates: 
$j$=1 (x y z),  $j$=2 (1-x y -z+1/2),  $j$=3 (-x+1/2 -y+1/2 1-z),  $j$=4 (x 1-y z+1/2). The ions generated by the C-lattice translation have the same magnetic moment
as the generating atom and are therefore omitted. The u,v,w coefficients are fixed by the symmetry for one set, but are independent for different sets. 
\label{table1}}
\begin{ruledtabular}
\begin{tabular}{lclclclc|c}
Ion&$\Gamma_1$&$\Gamma_2$&$\Gamma_3$&$\Gamma_4$\\
\multicolumn{5}{c}{4 a}\\
Co(41)&u v w& - &u v w& -\\
Co(43)&-u v -w& - &u -v w& -\\
 $i$=1,3&\multicolumn{4}{c}{8 f}\\
Co(i1)&u v w&u v w&u v w&u v w\\
Co(i2)&-u v -w&-u v -w&u -v w&u -v w\\
Co(i3)&u v w&-u -v -w&u v w&-u -v -w\\
Co(i4)&-u v -w&u -v w&u -v w&-u v -w\\
\end{tabular}
\end{ruledtabular}
\end{table}

\begin{figure}
\includegraphics[width=0.5\textwidth]{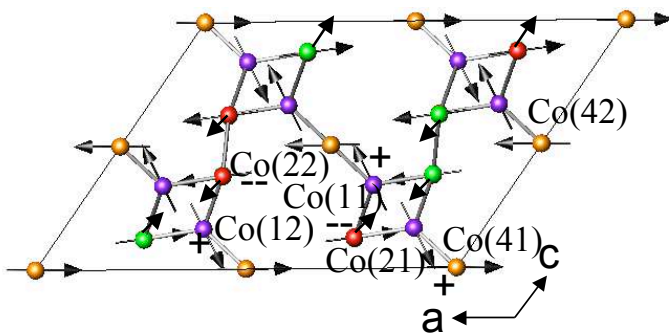}
\caption {(Color online) The $ac$-projection of the  low-temperature (5 K) magnetic structure of $\Co$. The four crystallographic sets of Co$^{2+}$ ions are shown by violet - Co(1), red - Co(2), green - Co(3) and orange - Co(4). The Co(21) and Co(32)  ions, as well as Co(22) and Co(32), superimpose on this projection. The pairs Co($ij$)/Co($ij$+1) have opposite M$_x$ and M$_z$ magnetic components, but the same M$_y$ component. The sign of M$_y$ is shown near the symbol of each ion. Spatial orientations of local magnetic moments are illustrated by three-dimensional vectors.}
\label{model}
\end{figure}

Refinement of the models has been performed with Fullprof~\cite{Rodriguez1993}.
The best agreement with experimental data ($R_M=6.6 \%$, Fig.~\ref{k0}) is obtained for a 3-dimensional canted $\Gamma_1$ model presented in Table~\ref{table2} and Fig.~\ref{model}. The $a^*bc$ orthogonal coordinate system defined in Inset to Fig.~\ref{k0} has been used. The Co moments reach the values of 4.21(7) $\mu_{B}$/Co(1), 4.4(1) $\mu_{B}$/Co(2), 3.8(1) $\mu_{B}$/Co(3) and 4.5(1) $\mu_{B}$/Co(4) at 5 K. These values are larger than the spin only component (3 $\mu_{B}$) of the Co$^{2+}$ ion confirming incomplete quenching of the orbital moment for all cobalt ions in this compound~\cite{Becker2006}.

%
\begin{table}
\caption{Refined parameters for the ${\bf{k}_2}$= 0 magnetic structure. M[$ \mu_{B}$] is the ordered magnetic moment,
with the M$_{xyz}$ components defined in the $a^*bc$ orthogonal coordinate system. The Co(1-4) sets are represented by the ions $j$=1 (x y z) and $j$=2 (1-x y -z+1/2). $\alpha$ [deg] is the canting angle between the magnetic moments of the 1st and the following ions.
\label{table2}}
\begin{ruledtabular}
\begin{tabular}{cccccc}
Ions&M$_x$&M$_y$&M$_z$&$\mu_{B}$/Co&$\alpha_{1i}$\\
  Co(11)&2.64(9)& 1.86(6)& 2.77( 9)&4.25(7)& 0\\
  Co(12)&-2.64(9)& 1.86(6)&-2.77( 9)&4.25(7)&125(2)\\
  Co(21)&-3.08(7)&-1.1( 1)& 3.08(7)&4.4(1)&95(2)\\
  Co(22)& 3.08(7)&-1.1( 1)&-3.08(7)&4.4(1)&97(2)\\
  Co(31)&-0.43(8)&-3.1( 1)& 2.22(8)&3.8(1)&93(2)\\
  Co(32)& 0.43(8)&-3.1( 1)&-2.22(8)&3.8(1)&131(2)\\
  Co(41)&-3.51(8)& 1.33( 9)& 2.48(8)&4.5(1)&90(2)\\
  Co(42)& 3.51(8)& 1.33( 9)&-2.48(8)&4.5(1)&75(2)\\
\end{tabular}
\end{ruledtabular}
\end{table}
%

%
%
The $\Co$ system could be identified as a canted antiferromagnet (weak ferromagnet) -- the angle between moments of the similar magnitude is smaller than 180$^{\circ}$, giving rise to a ferromagnetic component.
Each Co set ($i = 1,...,4$) has a different magnitude of the $M_y$ component. Ferromagnetic components associated with the Co(1) and Co(4) sets point in one direction, while those associated with Co(2) and Co(3) point in the opposite direction. The net moment along the $b$ axis given by the sum of the $M_y$ values in Table~\ref{table2} amounts to $\approx 0.5 \mu _B /Co$, which is in close agreement with the remanent moment obtained from the magnetization measurements (see, section $D$). As elaborated in the discussion section, the complex non-collinear canted magnetic structure of $\Co$ (Fig.~\ref{model}) stems from different involved interactions, first of all from the competing single ion anisotropy and exchange interactions.

\subsection{Magnetic susceptibility}

\subsubsection{High-temperature dependence}

%
%
\begin{figure}
\includegraphics[width=0.53\textwidth]{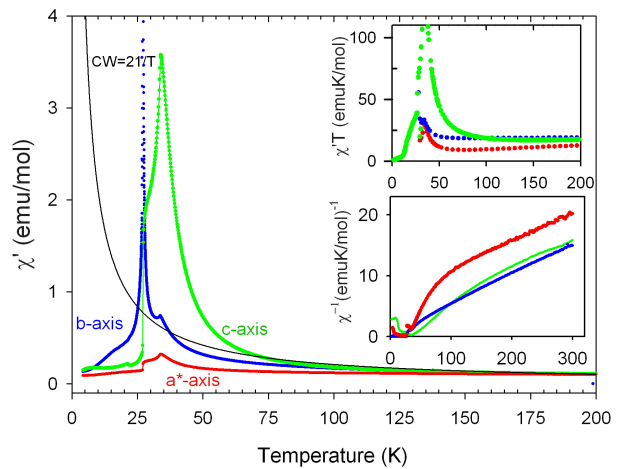}
\caption{(Color online) Main Panel: Temperature dependence of real component of ac susceptibility (3 Oe, 430 Hz) for measuring field oriented along three principal axes. DC susceptibility in small applied field (100 Oe) provides almost identical result. Black curve represents the Curie plot for free $S=3/2$, g=2.5, ions (Curie constant C=20.5 emuK/mol)  
Inset Top: Product $\chi_{DC} \cdot T$ vs. temperature. Inset Bottom: Susceptibility inverse $\chi_{DC}^{-1}$ vs. temperature. }
\label{dc-T}
\end{figure}

Results of ac and dc susceptibility measurements in three crystallographic directions $\astar$, $b$ and $c$ are shown in Fig.~\ref{dc-T}. Extending the original susceptibility report~\cite{Becker2006} these results document pronounced susceptibility anisotropy characterizing $\Co$ in a broad temperature range. For all three sample orientations the results reveal the presence of magnetic transition at $T_N = 34$ K, while the measurement along one axis only ($b$ axis) documents the presence of an additional, very pronounced transition at $T_C = 27$ K, in agreement with the original study. The present study shows that in the paramagnetic range the particular directional susceptibilities are remarkably different, the difference extending far above the transition temperature range. Unlike very anisotropic susceptibility the effective $g$-factor value, as determined from the high-temperature limit of the Curie-Weiss (CW) plot, was found to be pretty isotropic in the room temperature range. The values for $g$ were found to be very close to $g = 2.5$ (with S=3/2) for all three sample-to-field orientations, in full accordance with the original powder-sample data~\cite{Becker2006}. The value of the Weiss parameter $\theta $ of the CW plot varies depending on the chosen axis. We elaborate below that the behavior of the susceptibility is dominantly influenced by the single-ion anisotropy so the values of $\theta $ cannot be used to estimate the type and the strength of the involved exchange interactions. From the experimental side we note that there were only marginal sample-to-sample and batch-to-batch variations: The reported results thus rely only on the intrinsic crystal structure of the compound.

\begin{figure}
\includegraphics[width=0.4\textwidth]{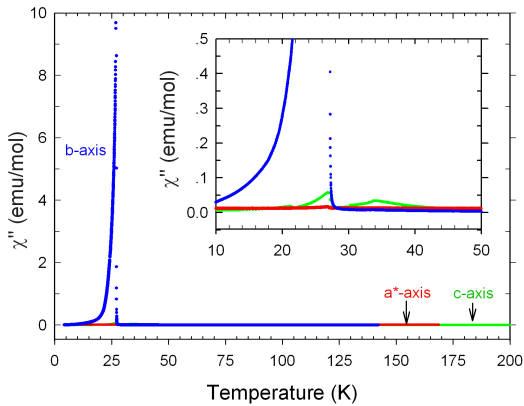}
\caption{(Color online) Temperature dependence of out-of-phase (imaginary) component of ac susceptibility for three principal axes. Above the transition range there is overlap of the data points for all three measurement directions. The data for $\chi'$ (Fig.~\ref{dc-T}) and $\chi''$ were taken within the same experimental runs. Inset: Zoomed transition temperature range. Residual signal in measurement along c-axis (green symbols) is ascribed to imperfect sample alignment and/or non-vanishing cross-talk between the in- and out-of-phase components in the phase sensitive detection.}
\label{imsus}
\end{figure}

Temperature dependence of the imaginary component of susceptibility is shown in Fig.~\ref{imsus}. These data are striking in two aspects. Firstly, imaginary susceptibility signal is present in the direction of the $b$-axis only. Secondly, the size and the sharpness of the imaginary susceptibility peak is, to the best of our knowledge, not observed in other magnetically ordered systems: under particular conditions (see Fig.~\ref{imsus}) imaginary peak is more than two times bigger than the peak in real susceptibility. Traditionally, peak in imaginary susceptibility is ascribed to dissipative ferromagnetic-domain dynamics, setting in simultaneously with the formation of the domains immediately below $T_c$. Due to its extraordinary properties imaginary susceptibility of $\Co$ will be subject of a separate publication.

\subsubsection{Susceptibility in magnetically ordered phase}

At $T_N = 34$ K a long-range magnetic order sets in. Both magnetic susceptibility and neutron diffraction data are consistent with antiferromagnetic ordering. Relative to the susceptibility maxima at $T_N$ there is a rapid drop of both $c-$ and $\astar-$ axis susceptibility by cooling below $T_N$ (Fig.~\ref{dc-T}), while $b-$ axis susceptibility simultaneously builds up. These findings are incompatible with simple uniaxial magnetic AF, which obviously does not take place in $\Co$. Three-dimensionally-canted and incommensurate order, documented by neutron diffraction results presented above, is fully compatible with susceptibility data. One notes that at $T_C$, marked by a very sharp peak of $b-$ axis susceptibility, $c-$ and $\astar-$ axis susceptibility exhibit a sudden drop (Fig.~\ref{dc-T}). The most natural explanation is that at $T_C$ an abrupt magnetic moment reorientation takes place such that a growing ferromagnetic component builds up at the expense of magnetic moments participating in the ICM ordering at higher temperatures. In accordance with the low-temperature magnetic structure (Fig.~\ref{model}) the latter observation shows that in magnetic ordering all magnetic degrees of freedom participate cooperatively, i.e., even if there exist separate magnetic fractions they contribute in mutual accord into the total susceptibility. One has to point out that low-field susceptibility studies reveal a remarkable feature of the system that one could not figure out from the crystallographic or low-temperature magnetic structures alone:  in spite of complexity of magnetic structure susceptibility data shows that there is a distinct axis, aligned approximately along $c-$ axis, which defines the direction of preferable spin orientation~\cite{fus0}. (A more precise determination of the preferred axis orientation is presented in Subsection \ref{MH}.) A fact that in the low-temperature magnetic structure the spins are aligned along very different directions, thus not along $c$- nor any other axis, is a consequence of different competing interactions ruling the spin geometry in the ground state.

\subsection{Anisotropic magnetic properties in Angular Overlap Model calculations}

Angular Overlap Model (AOM) calculations were performed to estimate the single ion magnetic anisotropy arising from the crystallographically inequivalent cobalt(II) centres.  AOM parameters for the Co-O and Co-Br bonding interactions were estimated from values of the ligand field splitting parameters tabulated for octahedral Co(H$_2$O)$_{6}^{2+}$ and tetrahedral CoBr$_{4}^{2-}$ complexes~\cite{Srivatsa1991,Cotton1961} assuming $e_\pi = 0.2e_\sigma$.  The parameter $e_\sigma$ was assumed to vary with distance \cite{Tregenna03} as a function of $1/r^5$ and $e_\pi$ as a function of $1/r^6$. The Racah and spin-orbit coupling parameters were fixed at 80 \% of their free-ion values, and the orbital Zeeman interaction reduced accordingly. The AOM matrices were constructed using LIGFIELD~\cite{ligfield} with the AOM parameters and angular coordinates calculated from structural data as input, employing all 120 functions of the $3d^7$ electronic configuration.  For the calculation of magnetic susceptibility curves, the AOM matrices were imported into the program MagProp\cite{TregennaNist}.
	
The magnetic moment per ion was calculated from the expression

\begin{equation}
\label{mag}
M _{ion}= \frac{\sum _{n} \Bigl( -\frac {dE _{n}}{dB} \Bigr) exp \Bigl( -\frac {E _{n}}{k _{B} T}\Bigr)}
{\sum _{n} exp\Bigl( -\frac {E _{n}}{k _{B} T} \Bigr) },			
\end{equation}
and the paramagnetic molar susceptibility from
\begin{equation}
\label{sus}
\chi _{p}=N_{A}\frac{M _{ion}}{B}.
\end{equation}
 
In these equations $B$ designates the external magnetic field, $k_B$ the Bolzmann constant and $N_A$ Avagadro's number.  The sum is over the n eigenstates of the Hamiltonian whose energies are designated by the symbol $E_n$.  The derivative in equation (1) was found according to the Hellman-Feynman theorem,

\begin{equation}
\label{HellFey}
 \frac {dE _{n}}{dB}=\langle \psi _n \mid \frac {d \hat {H}}{dB_0} \mid \psi _n \rangle.	
\end{equation}

The onset of short-range ferromagnetic order was incorporated by expressing the susceptibility as
\begin{equation}
\label{FerroInt}
\frac {1}{\chi _{M}}=\frac{1}{\chi _{p}}- \lambda ,
 \end{equation}
where $\lambda$ is the molecular field parameter.  

\begin{figure}
\includegraphics[width=0.5\textwidth]{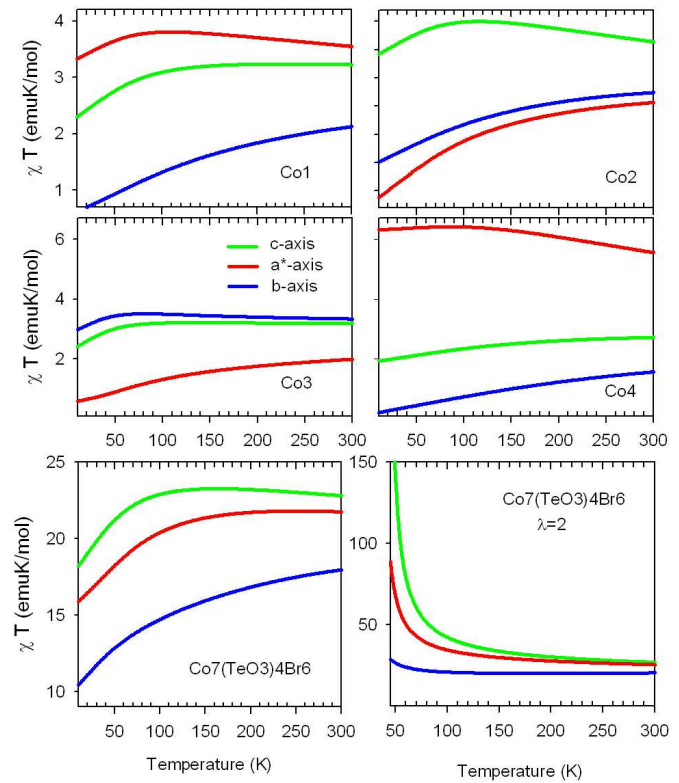}
\caption {Calculated products $\chi \cdot T$ for each of the four Co(i) octahedra employing full single-ion Hamiltonian and AOM. Calculated results for real single crystal as a whole is shown at bottom (plot at left). Results corrected for weak ferromagnetic interactions ($\lambda=2$, see text) is also shown, (plot at bottom right), to be compared with the experimental results for $\chi \cdot T$, Fig.\ref{dc-T}.}
\label{AOM-p}
\end{figure}

It is seen from Fig.\ref{AOM-p} that both experiment and AOM-based susceptibility calculations identify the component of the susceptibility tensor along the c-axis to be larger than the components along the a*- and b-axes. This would suggest that single-ion anisotropy indeed plays a major role in determining the observed magnetic anisotropy.  Note, however, that the calculated relative magnitudes of the susceptibility tensor along the a*- and b-axes are not in accordance with experiment. This could arise from our rather crude estimate of the AOM bonding parameters or may reflect the fact that no magnetic interaction paths have been introduced into the model.  In upgrading the model with appropriate interaction paths it would be natural to assume that the intralayer bc-plane exchange coupling dominates over the coupling along the out-of-plane a*-axis direction. Such corrections would certainly make the results of the model calculations in closer agreement with experiment.

As a result of the low-symmetry ligand field and spin-orbit coupling the $^{4}$T$_{1g}$(O$_{h}$) ground term is split into 6 Kramers doublets that are well-separated in energy.  In particular the first-excited Kramers doublet lies between 100 and 300 cm$^{-1}$ above the ground state for the four crystallographically inequivalent cobalt centres.  At sufficiently low temperatures, therefore, the electronic structure may be approximated as a pseudo- S=1/2 system, this being an approximation that is commonly employed for octahedrally co-ordinated cobalt(II) complexes\cite{Mabbs92}.

The $g^2$ tensor was calculated according to a method described in detail previously \cite{Scheifele08} in which the energies of the states of the lowest lying Kramers doublet are modeled by the eigenvalues of the S=1/2 spin-Hamiltonian,

 \begin{equation}
\label{spinH}
\hat H _s= \beta \bf{B} \cdot g \cdot S .
 \end{equation}

\begin{figure}
\includegraphics[width=0.5\textwidth]{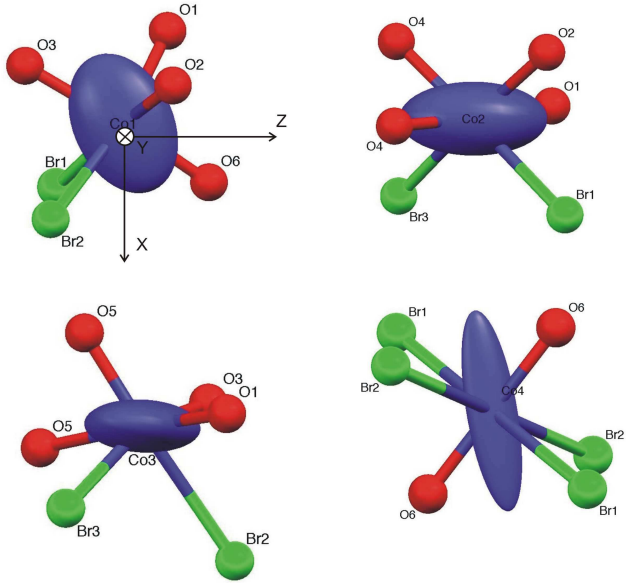}
\caption {The alignment of $g^2$-tensor (violet ellipsoid) and the ligands (oxygen in red, bromine in green) of four different Co set in the XYZ=$a^*bc$ orthogonal coordinate system.}
\label{AOM}
\end{figure}

The orientation of the $g^2$ -tensor in the a*bc reference frame and the ligand arrangement of four Co$^{2+}$ sets are shown in Fig.\ref{AOM}.

%
%

%
\begin{table}
\caption{Calculated components of the $\bf g$-matrix in the $a^*bc$  
reference frame and $\bf g'$
in the eigen coordinate frame.
\label{table3}}
\begin{ruledtabular}
\begin{tabular}{ccccccc}
$g_{ij}$&Co(1)&\multicolumn{2}{c}{$j$=1,..,3}&Co(2)&\multicolumn{2}{c} 
{$j$=1,..,3}\\
$i$=1&5.2251 & 1.0467  & 2.4724&2.8165  & 0.1922 & -0.1903\\
     2&1.0468 & 2.1070  & 0.5217&0.1922 &  3.4392  & 1.8252\\
     3&2.4724 & 0.5217  & 4.1107&-0.1903  & 1.8252 &  5.6682\\
$g'_{jj}$&2.2740&1.7214&7.4476&2.9437&2.2866&6.6936\\
$g_{ij}$&Co(3)&\multicolumn{2}{c}{$j$=1,..,3}&Co(4)&\multicolumn{2}{c} 
{$j$=1,..,3}\\
$i$=1&1.9333&  -1.0734&   0.8493&3.7126 &  0.8268 & -7.2514\\
     2& -1.0734&   4.8725&  -2.2980&0.1097 &  0.0012 & -1.2788\\
     3&0.8493&  -2.2980&   4.2329&-1.8364 & -0.9816  & 3.9631\\
$g'_{jj}$&2.2383&1.5742&7.2262&0.4946&-0.4940&9.3942\\
\end{tabular}
\end{ruledtabular}
\end{table}
%

A sizable c-axis susceptibility growth over the Curie margin (Fig.~\ref{dc-T}) is interpreted therefore as a combined effect of single-ion anisotropy and short range ferromagnetic correlations. On one side, it would be reasonable to assume that these correlations rely on weak ferromagnetism that, in principle, could accompany growth of the short-range antiferromagnetic order as the temperature approaches $T_N$ from above. This possibility has to be abandoned, however, as there is no trace of any ferromagnetic order (in the form of magnetic hysteresis, peak in imaginary susceptibility, see. Fig.~\ref{imsus}) setting in at, or immediately below, $T_N$. Ferromagnetic order, setting in below $T_C$, can only partially be related to high-temperature ferromagnetic correlations. As shown below, low-temperature ferromagnetism is manifested along b-axis only (thus not along the preferred c-axis) and its evolution is primarily correlated with the transformation/reorientation of the preformed ICM matrix.

\subsection{M-H studies}
\label{MH}

The most common experimental hallmarks of magnetically ordered ferromagnets and antiferromagnets are hysteris and spin-flop transition, respectively, known to characterize the respective magnetization vs. field (M-H) characteristics. Here we present the results of comprehensive M-H studies on $\Co$ demonstrating presence of \emph{both} of the mentioned hallmarks: which one of the two gets activated depends on the chosen orientation of principal axes with respect to the applied dc field. In brief, the field component along preferred axis, as imposed by single ion anistropy energy (approximately $c$-axis), activates antiferromagnetic response while the field component along $b$-axis activates ferromagnetic response. 

\subsubsection{Antiferromagnetic response}

\begin{figure}
\includegraphics[width=0.5\textwidth]{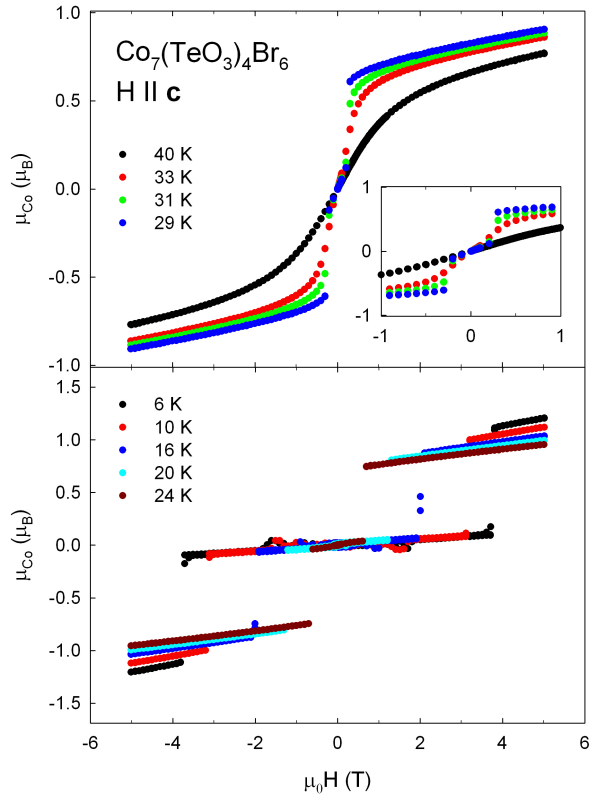}
\caption{(Color online) Magnetization (per one Co ion) vs. field characteristics for $c-$ axis direction in a broad temperature range. M(H) curves reveal almost switching but perfectly reversible behavior of magnetization in magnetically ordered phase.} 
\label{M-H_c}
\end{figure}

In Fig.~\ref{M-H_c} we first show the $M-H$ scans for the dc field applied parallel to the effective preferred axis, $c$-axis. For $T > T_N$ the $M-H$ curves are closed (i.e., do not show up hysteretic loops) and reveal no saturation for high fields indicating a paramagnetic state of the system.  For $T_C < T < T_N$ a sharp magnetization jump shows up around 20kOe. Magnitude of the magnetization jumps increases as $T$ approaches $T_C$. The field value at which a jump occurs, $H_{SF}^{c}$, does not change within this temperature interval. For $T < T_C$,  the magnetization transition additionally sharpen and $H_{SF}^{c}$ starts to grow by cooling and reaching $H_{SF}^{c}$=40 kOe at 5 K. Magnetization jumps are naturally ascribed to the spin-flop type of spin reorientation. The transition region is itself very narrow - at 20 K the transition is about 200 Oe wide.

%

\begin{figure}
\includegraphics[width=0.4\textwidth]{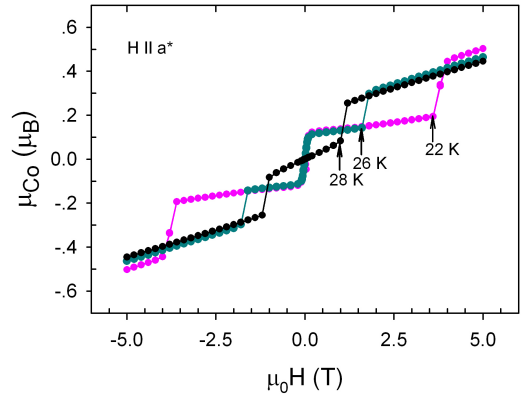}
\caption{(Color online) Magnetization (per one Co ion) vs. field curves for $\astar$ direction at three characteristic temperatures. Spin-flop transitions are marked with arrows.}
\label{M-H_a*}
\end{figure}

With the field applied along $\astar$-axis (Fig.~\ref{M-H_a*}) a very  similar behavior to the case of $H \parallel c$ has been obtained. In Fig.~\ref{M-H_a*} we show just several characteristic $M-H$ scans (out of many measured) taken at temperatures below and immediately above $T_c$. 

Qualitatively, the $M-H$ characteristics for $H \parallel c, \astar$ can be analyzed as a switching/spin reorientation (spin-flop) phenomenon  superimposed over some field-dependent background. The background is attributed to coherent rotation of sublatices' magnetization and, below $T_c$, to contributions from the $b$-axis ferromagnetism (s., next section). The background is more pronounced in the case $H \parallel \astar$ and one notes a drastic change of the background slope by cooling below $T_c$, Fig.~\ref{M-H_a*}. A rather pronounced background present in $H \parallel \astar$ orientation is a consequence of imperfect crystal alignment allowing a mixing-in of ferromagnetic $H \parallel b$ component. Otherwise, the major difference between the $M-H$ characteristics measured along $c$- and $\astar$- axis are about factor of 3-4 bigger spin-flop fields $H_{SF}^{\astar}$, in comparison with $H_{SF}^{c}$ at the same temperatures. The difference is attributed to geometrical reasons: in spin-flop transition the effective field component is only the one which is aligned along the preferred axis \cite{fus1}. From the spin-flop field values measured at 20 K ($H_{SF}^{\astar} = 4.7$ T and $H_{SF}^{c} = 1.3$ T) and crystal axes geometry (Inset to Fig.~\ref{k0}) one easily determines that the axis compatible with the minimum spin-flop field closes in the ($\astar,c$)-plane the angle $\phi$ ($\tan \phi = \frac{H_{SF}^{c}}{H_{SF}^{\astar}}$) with the $c$-axis. From the latter observation one concludes that the effective preffered axis is actually declined from the $c$-axis for the angle $\phi (\approx 15^0)$.

The observed sharpness of the spin-flop transition (Fig.~\ref{M-H_c}) represents a direct consequence of a sizable magnetic anisotropy characterizing the system. Generally, in a spin flop transition the spins of the sublatices rotate at $H_{SF}$ to the direction perpendicular to the field (and the preferred axis) direction. By subsequent field increase spins are coherently rotated to become aligned with the field only at the field value $H_{c}$. At $T=0 K$, the ratio of the two critical fields is expected to obey the relation~\cite{Carlin1977} $\frac{H_{c}(0)}{H_{SF}(0)}= (2\frac{H_{E}}{H_{A}}-1)^{1/2}$. Here, $H_{E}$ and $H_{A}$ represents the mean-field exchange field and the anisotropy field, respectively. Hence, in cases with inherently big anisotropy, such that $H_{A}$ approaches $H_{E}$, $H_{c}$ is not much bigger than $H_{SF}$ implying almost direct reorientation of a sublatice spin into the direction of preferred (i.e., magnetic field) axis. Such a transition is usually referred to as \emph {spin flip} and represents a generic feature of metamagnets~\cite{Carlin1977}. $\Co$ may therefore be considered as a typical metamagnet system. 

\begin{figure}
\includegraphics[width=0.5\textwidth]{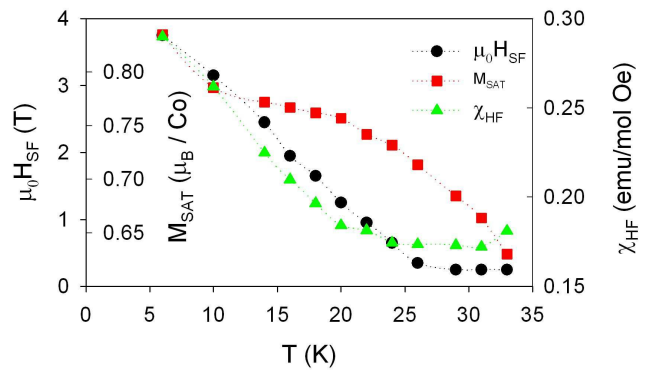}
\caption{(Color online) Temperature dependence of spin-flip field $H_{SF}^{c}$, high
field saturation magnetization $M_{sat}$ and the slope in $M–H$
curve at high fields $\chi _{HF}$, all for magnetic field applied along $c$-axis.}
\label{SpinFlopVsT}
\end{figure}

Temperature dependence of the field $H_{SF}$ is shown in Fig.~\ref{SpinFlopVsT}. Although there is no known generic analytic form of $H_{SF}(T)$, for metamagnets and antiferromagnets in general, the observed approximately linear temperature dependence is not very common~\cite{Carlin1977}. We note however that in the case of $\Co$ spin flip transition takes place within the magnetic structure which involves a ferromagnetically ordered component (along b-axis, next section) thus deviation from behavior known for other antiferromagnets/metamagnets should not be any surprising. Also, we point out that, in view of complex three-dimensional magnetic structure (Fig.~\ref{model}) the term `spin flip' cannot be applied literary in its text-book meaning, i.e.,  to mimic complete reversal of pairs originally anti-parallel oriented spins into the direction of preferred axis. From the magnitude of magnetization jump one cannot associate spin-flipping to any particular ion pair belonging to the two sublatices, which could perform, hypothetically, spin reversal isolated from the rest of structure. Instead, it's more probable that in the energy landscape of magnetic structure as a whole there are two neighboring minima becoming equal in energy by the application of magnetic field $H_{SF}$ along the effective preferred axis.

%
%

%

\subsubsection{Ferromagnetic response}

\begin{figure}
\includegraphics[width=0.45\textwidth]{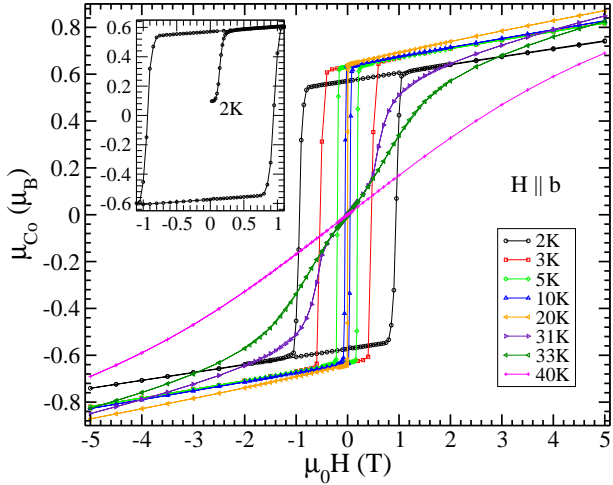}
\caption{(Color online) Main Panel: Magnetic hysteresis loops for the $b$-axis direction at different temperatures. Magnetization is scaled to one Co ion. Inset: Virgin hysteresis curve at 2 K. At higher temperatures there is a more abrupt magnetization transition to quasi-saturation by the application of magnetic field of the order of 100 Oe or less.}
\label{M-H_b}
\end{figure}

When the magnetic field is applied along the $b$ axis, along which the ferromagnetic component exists, remarkably different behavior is observed, Fig.~\ref{M-H_b}. Below $T_N$ a small kink develops in the $M-H$ curves but contrary to the $H \parallel \astar,c$ cases it shifts towards lower values as temperature is decreased. On further cooling below $T_C$ a narrow, almost rectangular hysteresis opens up around zero, indicating a formation of ferromagnetic domains. Initial magnetization also shows ferromagnetic character, achieving the saturation value in the virgin curve by applying only 100 Oe at 10 K.

\begin{figure}
\includegraphics[width=0.45\textwidth]{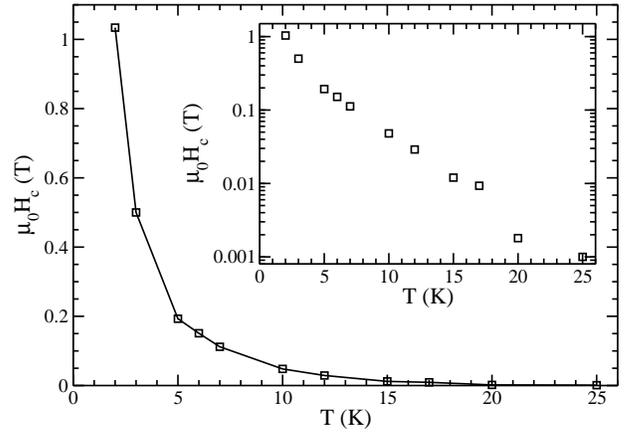}
\caption{Temperature dependence of the coercive field for b-axis field direction. Solid line is guide for the eye only.}
\label{coercive}
\end{figure}

Two characteristic features of a hysteresis loop are remanence and coercivity. Below $T_C$ the remanent moment is practically constant with a value $\mu _{rem} \approx 0.625 \mu _B$ (per one Co ion). On the other hand, the coercive field shows a linear dependence on temperature in the log-lin plot, as indicated in the inset of Fig.~\ref{coercive}. The relation

%
\begin{equation}
\label{exp-coercive}
H_C (T) = H_C^0 \cdot e^{-\alpha T}
\end{equation}
has been found to describe well the behavior of many nanostructured magnetic systems, like thin magnetic films~\cite{Vertesy1998} and amorphous systems~\cite{Ribas1995,Read1984,Cresswell1990,Pajic2007}. The common feature in those systems was the presence of magnetic clusters with the well defined anisotropy barrier, where jumps of magnetic moments of the clusters over the barriers are temperature assisted.

The prominent feature of ferromagnetic order in $\Co$ is a very sharp transitions between the two saturation states, giving rise to almost rectangular hysteresis. Very often, rectangular hysteresis is observed in multilayers~\cite{Nakajima1993,Weller2001} where magnetic and nonmagnetic layers are stacked on top of each other (Co and Pt for example~\cite{Weller2001}). Besides the choice of the constituent materials, magnetic properties of multilayers depend strongly on thickness of the individual layers, as well as on the growth process. A sudden reversal of magnetization occurs when for a critical field one nucleation site for a reversed domain is generated and the avalanche effect is propagated throughout the material via strong exchange coupling. To the best of our knowledge, $\Co$ is the first `non-multilayer' material exhibiting the effect of rectangular hysteresis. In view of its layered structure (stack of bc-plane layers), embedding the ferromagnetic component within the layers, the latter interpretation seems at least as a consistent possibility.

Alternatively, the rectangular hysteresis might actually be underlined by Stoner-Wohlfarth single-domain model~\cite{Morrish2001,Blundell2001}. Our single-crystalline $\Co$ samples are certainly to big to represent a monodomain below $T_C$, as one easily verifies from the fact that below $T_C$ the net magnetization along $b$-axis (as well as along any other axis) in $H_{dc}=0$ is $M=0$. From the initial (virgin) curve (Fig.~\ref{M-H_b}) one notes however that a quasi-saturation is achieved already in a very small applied field rendering the sample practically monodomain in any bigger fields. For this reason consideration of $M-H$ hysteresis in terms of Stoner-Wohlfarth model makes sense. In the latter model the total magnetic energy consists of the two terms, the energy of uniform magnetization in the field sweeping up and down and the uniaxial anisotropy energy of magnetization $M$ as it gets deflected from the preferred axis. Total energy minimization results with the hysteretic $M-H$ curves, the shapes of which are precisely determined by the direction of applied field with respect to magnetization/preferred axis~\cite{Morrish2001}. In the geometry of magnetic field aligned with the magnetization, being realized in this study (Fig.~\ref{M-H_b}), Stoner-Wohlfarth model generates strictly rectangular hysteresis loops characterized by coercive field equal to anisotropy field $H_C=2K/M$ ($K$ is a constant of uniaxial anisotropy). At this field value the energy of magnetization in ramping applied field just overcomes the anisotropy energy enabling spin reversal to take place. 

%

In the particular case of $\Co$ it is interesting to note that the effective preferred axis is oriented approximately along $c$-axis, not along the magnetization axis ($b$-axis in our case) as is the case in the original formulation of Stoner-Wohlfarth model. In the case of $\Co$ one has to bear in mind that ferromagnetic moment is just a component of canted three-dimensional order thus representing, as pointed out in this article's title, a byproduct of global antiferromagnetism. It is also interesting to compare values of the two critical fields $H_{SF}$ and $H_C$  at the same temperatures. $H_{SF}$ is systematically bigger than $H_C$ for, approximately, an order of magnitude. This finding is consistent with elementary understanding of dynamics of magnetically ordered systems: To manipulate ferromagnetically ordered spins one has to apply field at the order of anisotropy field $H_A$ while to do the same with antiferromagnetically ordered spins one has to apply much larger, exchange-enhanced field $(H_A H_E)^{1/2}$ (one bears in mind that $H_E \gg H_A$). The latter aspect has recently been pointed out \cite{Kimel2004} in the context of requirements for ultrafast spin reorientation technologies.

%
%
%
%
%

\section{Discussion}
\label{Discussion}

In order to discus the results let us first recapitulate the main observations presented herewith. 

At $T_N$, a long-range magnetic ordering of $\Co$ takes place, with a temperature dependent and incommensurate propagation wave vector\cite{fus2} $k_{ICM}$.  In direct space the latter temperature dependence might be associated with global spin  reorientations giving rise to ferromagnetic component along the $b$-axis, setting in at $T_C$. Antiferromagnetic response along the two other perpendicular axes ($\astar$- and $c$-axis) is kept unchanged, however. Upon lowering the temperature below $T_C$, the commensurate (CM) structure is stabilized and  coexists with the ICM structure in the short temperature interval. The antiferromagnetic backbone of the CM structure provides a natural explanation for the spin flop (or spin flip) transitions and rectangular-shaped hysteresis loops observed in sweeping the dc field along $\astar$,$c$-axis, and $b$ axis, respectively. When the first nucleation center for the reversed domain is created, the backbone structure becomes unstable relative to the 180$^0$ rotation of all the moments, reversing the total magnetization in an extremely narrow field interval ($\sim  100$ Oe at 20 K and $\sim 400$ Oe at 5 K).

In an attempt to interpret these observation one has to point out that low C2/c symmetry and 4 different (distorted) environments for magnetic ions make the modeling for $\Co$ extremely difficult. Instead of attempting the latter  here we just elaborate magnetic interactions and mechanism found responsible for complex magnetic ordering of  $\Co$. Most obviously, there are at least two equally important interactions ruling magnetism of this compound:  exchange interaction and single ion anisotropy energy. 

Exchange interactions provide a necessary framework for magnetic ordering to set in. As the order established at $T_N$ is incommensurate the exchange interactions $J_{ij}$ between magnetic moments of Co$^{2+}$ ions necessarily extends beyond nearest neighbors, differing in size as well in sign. Single ion anisotropy plays pronounced role in ruling anisotropic paramagnetic susceptibility. In the ordered phase it provides dominant, or at least sizable, contribution to macroscopic magnetocrystalline anisotropy, $K$. (Other possible contributions to $K$ rely on magnetic-dipolar anisotropy and exchange-interaction anisotropy, see, e.g. Ref.~\onlinecite{Besser1967}). In the ordered phase $K$ underlays the phenomenon of spin fl(o)ip, introducing the temperature through temperature dependence of magnetization~\cite{Zener1954}. 
 
Now, just on ground of specific exchange interaction network $J_{ij}$ compatible with the incommensurate order and the explicit temperature dependence of magnetocrystalline energy one is able to interpret, at least qualitatively, several important features of $\Co$ in the ordered state, like the temperature dependence of $k_{ICM}$ and a sudden appearance of ferromagnetism at $T_C$.

Depending on whether the exchange integrals $J_{ij}$ are considered as isotropic or anisotropic there are two possible scenarios. In the first, one notes a striking similarity of magnetic ordering patterns of $\Co$ and rare-earth metals Dysprosium (Dy) and Terbium (Tb). The latter elements acquire incommensurate-helical order at their particular $T_N$ featuring temperature dependent $k_{ICM}$ and ferromagnetic transition, taking place a few degrees below $T_N$.  By keeping aside itinerant-magnetism peculiarities, generally important for Dy and Tb, the ordering phenomenology itself can convincingly be interpreted~\cite{Nagamiya1967} on basis of an interplay between the isotropic exchange integrals and temperature dependent magnetocrystalline energy. Closer examination of the driving force responsible for ferromagnetism of Dy and Tb showed however that magnetostriction plays perhaps a more direct role than the temperature dependence of single ion anisotropy~\cite{Cooper1967}. Whether magnetostriction sets in at $T_c$ in $\Co$ as well is not fully resolved as yet. A high-resolution diffraction study is needed to detect magnetostriction-related structural changes in vicinity of $T_c$.

In the second scenario the exchange interaction is considered as being anisotropic, either due to directional dependence of exchange integrals or due to antisymmetric form of the related spin-spin operator in the interaction Hamiltonian. In the former case the scalars $J_{ij}$ are replaced by the appropriate tensor. Competition of anisotropic exchange with uniaxial single ion anisotropy, as analyzed in mean-field approximation~\cite{deNeef1974}, gives rise to different types of possible spin orders minimizing the Gibbs free energy, allowing for spin reorientations below corresponding temperatures\cite{Levison1969}. In the latter case anisotropic exchange interaction is manifested as an antisymmetric (Dzyaloshinsky-Morya, DM) spin-spin interaction. The DM interaction\cite{Moriya1960PRL} is directly responsible for numerous cases of systems revealing weak ferromagnetism in canted antiferromagnets\cite{Carlin1977,Blundell2001}. As ferromagnetism in $\Co$ is certainly of canted type (Fig.~\ref{model}), emerging abruptly from global antiferromagnetism, it is plausible to assume relevance of DM interaction, for $\Co$ as well. With this respect the mechanism ruling the spin reorientation in $\Co$ could be closely related to spontaneous spin-flip (Morin transition), taking place deep in antiferromagnetic phase of hematite, $\alpha-Fe_2O_3$, being accompanied by a DM-based ferromagnetic component. A driving force for the Morin transition is identified in competition of single-ion anisotropy with long-range dipolar anisotropy term~\cite{Artman65}.

As pointed out by Becker and coworkers~\cite{Becker2006}, $\Co$ is composed of chains running along the $b$ axis. The chains contain Co(2) and Co(3) ions and are interconnected with Co(1) ions to form the $bc$ layers. The layers are linked through the Co(1) - Co(4) - Co(1) connection. As indicated by AOM calculations, Co(1) and Co(4) ions are well described in the framework of single-ion anisotropy even in the low temperature limit. On the other hand, Co(2) and Co(3) ions seem to be substantially influenced by the exchange interactions through the ligands, indicating a good connection along the chain direction and rather weak perpendicular to them. The preliminary inelastic neutron scattering experiments along the $c$ direction indicate the presence of the dispersionless mode around 4 meV, pointing to a weak coupling along the $c$ direction, corroborating the AOM results.

%
%
%
%

\section{Conclusions}
\label{Conclusion}

In its ground state $\Co$ is a three-dimensional canted magnetically ordered system revealing antiferromagnetically compensated sublattices in all, but the $b$-axis direction. The low temperature magnetic structure is stabilized through temperature dependent incommensurate wave vector accompanied by an abrupt emergence of ferromagnetic component along $b$-axis. Magnetic susceptibility studies demonstrate extreme anisotropy characterizing $\Co$. Although the ferromagnetic component establishes strictly along the $b$-axis, the effective preferred axis, imposed by a sizable single ion anisotropy term of octahedrally coordinated Co$^{2+}$ ions, is directed approximately along $c$-axis. Hidden in a three-dimensionally canted spin arrangement the latter axis represents, as clearly shown in susceptibility and $M-H$ studies, a real `backbone' of antiferromagnetic order. Accordingly, magnetic field (or its component) has to be applied along this axis to induce entirely reversible spin flop transition. Closer examination of $M-H$ scans taken along $c$- and $\astar$- axis shows that the effective preferred axis is actually declined from the $c$-axis for an angle of approximately $15^0$. From the extreme sharpness of the transition it is concluded that a phenomenon of spin flip (instead of a spin flop) better describes the observations, classifying $\Co$ into the category of metamagnets. Ferromagnetic response, restricted to the direction of $b$-axis only, has been related to the phenomenology of multilayers and/or of the Stoner-Wohlfarth model due to strikingly rectangular hysteresis loops. $\Co$ obviously represents a remarkable magnetic system manifesting competition of various magnetic interactions. In the competition there is primarily a complex network of exchange interactions and a single ion anisotropy energy. Most probably the single ion anisotropy represents just one possible component of a more complex magnetocrystalline anisotropy, relevant for energy balance in the ordered phase, and in the article other alternatives are discussed along the lines of knowledge accumulated during the decades.

\section{Acknowledgments}

M.P., I. \v Z., and D.P. acknowledge financing from the projects 035-0352843-2845 and 119-1191458-1017 of the Croatian Ministry of Science, Education and Sport. M.P. thanks Djuro Drobac, Institute of Physics, Zagreb, for help in ac susceptibility measurements. D.P. is grateful to Kre\v so Zadro, Dept. of Physics, Faculty of Science,
University of Zagreb, for advices related to magnetization measurements. H.B. thanks the NCCR research pool MaNEP of the Swiss National Science Foundation for support in sample preparation. Neutron facilities of SINQ, Paul Scherrer Institute, Villigen, Switzerland are also gratefully acknowledged.


\begin{thebibliography}{99}
\bibitem{fus2}
Due to the strong neutron absorption cross section of Co ions and the specific, plate-like shape of the crystals, the determination of the low-symmetry ICM structure for $T_C < T < T_N$ was not successful.
\end{thebibliography}
\end{document}